\documentclass[11pt]{article}

\usepackage{xcolor}
\usepackage{titlesec}
\usepackage{abstract}
\usepackage{lipsum}
\usepackage{multicol}
\usepackage{tocloft}
\usepackage{graphicx}
\usepackage[footnotesize,hang,bf]{caption}
\usepackage[font=footnotesize]{subcaption}
\usepackage{times}
\usepackage{amssymb,amsmath}
\usepackage{footmisc}
\usepackage{appendix}
\usepackage{epstopdf}
\usepackage[leftcaption]{sidecap}
\usepackage{wrapfig}
\usepackage{enumerate}
\usepackage{fancyhdr}
\usepackage{makeidx}
\usepackage{indentfirst}
\usepackage{multirow}
\usepackage{color}
\usepackage{fancybox}
\usepackage{listing}
\usepackage[hyphens]{url}
\usepackage[intoc]{nomencl}
\usepackage{array}
\usepackage{cfxminted}

\newcommand{\tighterenum}{\setlength{\parskip}{-4pt}}

\definecolor{mtbash}{rgb}{0.98, 0.98, 0.98}
\definecolor{mtconsole}{rgb}{0.98, 0.98, 0.98}
\definecolor{mtcode}{rgb}{1.0, 1.0, 1.0}
\definecolor{mtfortran}{rgb}{1.0, 1.0, 1.0}

\titleformat{\section}{\raggedright\bfseries\large\scshape}{\thesection.}{0.5em}{}
\titleformat{\subsection}{\raggedright\normalfont\scshape}{\thesubsection.}{0.5em}{}
\titleformat{\subsubsection}{\raggedright\normalfont\small\scshape}{\thesubsubsection.}{0.5em}{}

\definecolor{linkcolor}{rgb}{0.00,0.00,0.75}
\definecolor{citecolor}{rgb}{0.00,0.00,0.75}
\definecolor{linkbordercolor}{rgb}{0.00,0.00,0.75}

\usepackage{hyperref}
\hypersetup{%
  linktocpage=true,
  colorlinks=true,
  urlcolor=linkcolor, 
  linkcolor=linkcolor,
  menucolor=linkcolor,
  pdfpagemode=UseOutlines,    
  citecolor=citecolor,
  linkbordercolor=linkbordercolor,
  citebordercolor=linkbordercolor,
  urlbordercolor=linkbordercolor,
      pdftitle={Cluster-Level Tuning of a Shallow Water Equation Solver on the Intel MIC Architecture},
      pdfsubject={White Paper},
      pdfauthor={Andrey Vladimirov and Cliff Addison},
      pdfkeywords={bandwidth, cache, CFD, communication, InfiniBand, Ivy Bridge, MIC, MPI, Mellanox, SIMD, Xeon Phi, optimization}
}
\makeatletter
\Hy@AtBeginDocument{%
  \def\@pdfborder{0 0 1}
  \def\@pdfborderstyle{/S/U/W 0.6}
}
\makeatother

\makeatletter
\def\PY@reset{\let\PY@it=\relax \let\PY@bf=\relax%
    \let\PY@ul=\relax \let\PY@tc=\relax%
    \let\PY@bc=\relax \let\PY@ff=\relax}
\def\PY@tok#1{\csname PY@tok@#1\endcsname}
\def\PY@toks#1+{\ifx\relax#1\empty\else%
    \PY@tok{#1}\expandafter\PY@toks\fi}
\def\PY@do#1{\PY@bc{\PY@tc{\PY@ul{%
    \PY@it{\PY@bf{\PY@ff{#1}}}}}}}
\def\PY#1#2{\PY@reset\PY@toks#1+\relax+\PY@do{#2}}

\expandafter\def\csname PY@tok@gd\endcsname{\def\PY@tc##1{\textcolor[rgb]{0.63,0.00,0.00}{##1}}}
\expandafter\def\csname PY@tok@gu\endcsname{\let\PY@bf=\textbf\def\PY@tc##1{\textcolor[rgb]{0.50,0.00,0.50}{##1}}}
\expandafter\def\csname PY@tok@gt\endcsname{\def\PY@tc##1{\textcolor[rgb]{0.00,0.27,0.87}{##1}}}
\expandafter\def\csname PY@tok@gs\endcsname{\let\PY@bf=\textbf}
\expandafter\def\csname PY@tok@gr\endcsname{\def\PY@tc##1{\textcolor[rgb]{1.00,0.00,0.00}{##1}}}
\expandafter\def\csname PY@tok@cm\endcsname{\let\PY@it=\textit\def\PY@tc##1{\textcolor[rgb]{0.25,0.50,0.50}{##1}}}
\expandafter\def\csname PY@tok@vg\endcsname{\def\PY@tc##1{\textcolor[rgb]{0.10,0.09,0.49}{##1}}}
\expandafter\def\csname PY@tok@m\endcsname{\def\PY@tc##1{\textcolor[rgb]{0.40,0.40,0.40}{##1}}}
\expandafter\def\csname PY@tok@mh\endcsname{\def\PY@tc##1{\textcolor[rgb]{0.40,0.40,0.40}{##1}}}
\expandafter\def\csname PY@tok@go\endcsname{\def\PY@tc##1{\textcolor[rgb]{0.53,0.53,0.53}{##1}}}
\expandafter\def\csname PY@tok@ge\endcsname{\let\PY@it=\textit}
\expandafter\def\csname PY@tok@vc\endcsname{\def\PY@tc##1{\textcolor[rgb]{0.10,0.09,0.49}{##1}}}
\expandafter\def\csname PY@tok@il\endcsname{\def\PY@tc##1{\textcolor[rgb]{0.40,0.40,0.40}{##1}}}
\expandafter\def\csname PY@tok@cs\endcsname{\let\PY@it=\textit\def\PY@tc##1{\textcolor[rgb]{0.25,0.50,0.50}{##1}}}
\expandafter\def\csname PY@tok@cp\endcsname{\def\PY@tc##1{\textcolor[rgb]{0.74,0.48,0.00}{##1}}}
\expandafter\def\csname PY@tok@gi\endcsname{\def\PY@tc##1{\textcolor[rgb]{0.00,0.63,0.00}{##1}}}
\expandafter\def\csname PY@tok@gh\endcsname{\let\PY@bf=\textbf\def\PY@tc##1{\textcolor[rgb]{0.00,0.00,0.50}{##1}}}
\expandafter\def\csname PY@tok@ni\endcsname{\let\PY@bf=\textbf\def\PY@tc##1{\textcolor[rgb]{0.60,0.60,0.60}{##1}}}
\expandafter\def\csname PY@tok@nl\endcsname{\def\PY@tc##1{\textcolor[rgb]{0.63,0.63,0.00}{##1}}}
\expandafter\def\csname PY@tok@nn\endcsname{\let\PY@bf=\textbf\def\PY@tc##1{\textcolor[rgb]{0.00,0.00,1.00}{##1}}}
\expandafter\def\csname PY@tok@no\endcsname{\def\PY@tc##1{\textcolor[rgb]{0.53,0.00,0.00}{##1}}}
\expandafter\def\csname PY@tok@na\endcsname{\def\PY@tc##1{\textcolor[rgb]{0.49,0.56,0.16}{##1}}}
\expandafter\def\csname PY@tok@nb\endcsname{\def\PY@tc##1{\textcolor[rgb]{0.00,0.50,0.00}{##1}}}
\expandafter\def\csname PY@tok@nc\endcsname{\let\PY@bf=\textbf\def\PY@tc##1{\textcolor[rgb]{0.00,0.00,1.00}{##1}}}
\expandafter\def\csname PY@tok@nd\endcsname{\def\PY@tc##1{\textcolor[rgb]{0.67,0.13,1.00}{##1}}}
\expandafter\def\csname PY@tok@ne\endcsname{\let\PY@bf=\textbf\def\PY@tc##1{\textcolor[rgb]{0.82,0.25,0.23}{##1}}}
\expandafter\def\csname PY@tok@nf\endcsname{\def\PY@tc##1{\textcolor[rgb]{0.00,0.00,1.00}{##1}}}
\expandafter\def\csname PY@tok@si\endcsname{\let\PY@bf=\textbf\def\PY@tc##1{\textcolor[rgb]{0.73,0.40,0.53}{##1}}}
\expandafter\def\csname PY@tok@s2\endcsname{\def\PY@tc##1{\textcolor[rgb]{0.73,0.13,0.13}{##1}}}
\expandafter\def\csname PY@tok@vi\endcsname{\def\PY@tc##1{\textcolor[rgb]{0.10,0.09,0.49}{##1}}}
\expandafter\def\csname PY@tok@nt\endcsname{\let\PY@bf=\textbf\def\PY@tc##1{\textcolor[rgb]{0.00,0.50,0.00}{##1}}}
\expandafter\def\csname PY@tok@nv\endcsname{\def\PY@tc##1{\textcolor[rgb]{0.10,0.09,0.49}{##1}}}
\expandafter\def\csname PY@tok@s1\endcsname{\def\PY@tc##1{\textcolor[rgb]{0.73,0.13,0.13}{##1}}}
\expandafter\def\csname PY@tok@sh\endcsname{\def\PY@tc##1{\textcolor[rgb]{0.73,0.13,0.13}{##1}}}
\expandafter\def\csname PY@tok@sc\endcsname{\def\PY@tc##1{\textcolor[rgb]{0.73,0.13,0.13}{##1}}}
\expandafter\def\csname PY@tok@sx\endcsname{\def\PY@tc##1{\textcolor[rgb]{0.00,0.50,0.00}{##1}}}
\expandafter\def\csname PY@tok@bp\endcsname{\def\PY@tc##1{\textcolor[rgb]{0.00,0.50,0.00}{##1}}}
\expandafter\def\csname PY@tok@c1\endcsname{\let\PY@it=\textit\def\PY@tc##1{\textcolor[rgb]{0.25,0.50,0.50}{##1}}}
\expandafter\def\csname PY@tok@kc\endcsname{\let\PY@bf=\textbf\def\PY@tc##1{\textcolor[rgb]{0.00,0.50,0.00}{##1}}}
\expandafter\def\csname PY@tok@c\endcsname{\let\PY@it=\textit\def\PY@tc##1{\textcolor[rgb]{0.25,0.50,0.50}{##1}}}
\expandafter\def\csname PY@tok@mf\endcsname{\def\PY@tc##1{\textcolor[rgb]{0.40,0.40,0.40}{##1}}}
\expandafter\def\csname PY@tok@err\endcsname{\def\PY@bc##1{\setlength{\fboxsep}{0pt}\fcolorbox[rgb]{1.00,0.00,0.00}{1,1,1}{\strut ##1}}}
\expandafter\def\csname PY@tok@kd\endcsname{\let\PY@bf=\textbf\def\PY@tc##1{\textcolor[rgb]{0.00,0.50,0.00}{##1}}}
\expandafter\def\csname PY@tok@ss\endcsname{\def\PY@tc##1{\textcolor[rgb]{0.10,0.09,0.49}{##1}}}
\expandafter\def\csname PY@tok@sr\endcsname{\def\PY@tc##1{\textcolor[rgb]{0.73,0.40,0.53}{##1}}}
\expandafter\def\csname PY@tok@mo\endcsname{\def\PY@tc##1{\textcolor[rgb]{0.40,0.40,0.40}{##1}}}
\expandafter\def\csname PY@tok@kn\endcsname{\let\PY@bf=\textbf\def\PY@tc##1{\textcolor[rgb]{0.00,0.50,0.00}{##1}}}
\expandafter\def\csname PY@tok@mi\endcsname{\def\PY@tc##1{\textcolor[rgb]{0.40,0.40,0.40}{##1}}}
\expandafter\def\csname PY@tok@gp\endcsname{\let\PY@bf=\textbf\def\PY@tc##1{\textcolor[rgb]{0.00,0.00,0.50}{##1}}}
\expandafter\def\csname PY@tok@o\endcsname{\def\PY@tc##1{\textcolor[rgb]{0.40,0.40,0.40}{##1}}}
\expandafter\def\csname PY@tok@kr\endcsname{\let\PY@bf=\textbf\def\PY@tc##1{\textcolor[rgb]{0.00,0.50,0.00}{##1}}}
\expandafter\def\csname PY@tok@s\endcsname{\def\PY@tc##1{\textcolor[rgb]{0.73,0.13,0.13}{##1}}}
\expandafter\def\csname PY@tok@kp\endcsname{\def\PY@tc##1{\textcolor[rgb]{0.00,0.50,0.00}{##1}}}
\expandafter\def\csname PY@tok@w\endcsname{\def\PY@tc##1{\textcolor[rgb]{0.73,0.73,0.73}{##1}}}
\expandafter\def\csname PY@tok@kt\endcsname{\def\PY@tc##1{\textcolor[rgb]{0.69,0.00,0.25}{##1}}}
\expandafter\def\csname PY@tok@ow\endcsname{\let\PY@bf=\textbf\def\PY@tc##1{\textcolor[rgb]{0.67,0.13,1.00}{##1}}}
\expandafter\def\csname PY@tok@sb\endcsname{\def\PY@tc##1{\textcolor[rgb]{0.73,0.13,0.13}{##1}}}
\expandafter\def\csname PY@tok@k\endcsname{\let\PY@bf=\textbf\def\PY@tc##1{\textcolor[rgb]{0.00,0.50,0.00}{##1}}}
\expandafter\def\csname PY@tok@se\endcsname{\let\PY@bf=\textbf\def\PY@tc##1{\textcolor[rgb]{0.73,0.40,0.13}{##1}}}
\expandafter\def\csname PY@tok@sd\endcsname{\let\PY@it=\textit\def\PY@tc##1{\textcolor[rgb]{0.73,0.13,0.13}{##1}}}

\def\PYZbs{\char`\\}
\def\PYZus{\char`\_}

\def\PYZam{\char`\&}

\def\PYZgt{\char`\>}

\def\PYZdl{\char`\$}
\def\PYZhy{\char`\-}

\def\PYZdq{\char`\"}

\makeatother

\definecolor{quotecolor}{RGB}{255,255,200}

\ifdefined\mobiledoc
 \newcommand{\opttwocolumn}{\onecolumn}
 
 \newcommand{\optcolstart}{}
 \newcommand{\optcolend}{}
 \newcommand{\optcolsep}{\newpage}
 \newcommand{\colfaxwidth}{2.8 in}
 \fancyfootoffset{-2pt}
\else
 \newcommand{\opttwocolumn}{\twocolumn}
 
 \newcommand{\optcolstart}{\begin{multicols}{2}}
 \newcommand{\optcolend}{\end{multicols}}
 \newcommand{\optcolsep}{\vfill\columnbreak}
 \newcommand{\colfaxwidth}{5.7 in}
\fi

\definecolor{llblue}{rgb}{0.93,0.93,1.0}
	
\makenomenclature

\begin{document}
\sloppy 

{\onecolumn

\author{\slshape Andrey Vladimirov$^1$ and Cliff Addison$^2$\\~\\ \small \slshape $^1$ Colfax International \\  \small \slshape $^2$ University of Liverpool}

\title{\Large\scshape
\vskip -2.5 em
Cluster-Level Tuning\\
of a Shallow Water Equation Solver\\
on the Intel MIC Architecture
}

\date{\small May 12, 2014}

{\linespread{0.8} \maketitle}

\optcolstart

\begin{abstract}
\footnotesize
The paper demonstrates the optimization of the execution environment 
of a hybrid OpenMP+MPI computational fluid dynamics code (shallow water equation solver)
on a cluster enabled with Intel Xeon Phi coprocessors.
The discussion includes:

\begin{enumerate}[1.]
\setlength{\parskip}{-1pt}
\item Controlling the number and affinity of OpenMP threads to optimize access to memory bandwidth;
\item Tuning the inter-operation of OpenMP and MPI to partition the problem for better data locality;
\item Ordering the MPI ranks in a way that directs some of the traffic into faster communication channels;
\item Using efficient peer-to-peer communication between Xeon Phi coprocessors based on the InfiniBand fabric.
\end{enumerate}

With tuning, the application has 90\% percent efficiency of parallel scaling 
up to 8 Intel Xeon Phi coprocessors in 2 compute nodes.
For larger problems, scalability is even better, because of the greater computation to communication ratio.
However, problems of that size do not fit in the memory of one coprocessor.

The performance of the solver on one Intel Xeon~Phi coprocessor~7120P exceeds the performance on a dual-socket \mbox{Intel~Xeon~E5-2697~v2} CPU by a factor of 1.6x. 
In a 2-node cluster with 4 coprocessors per compute node, the MIC architecture yields 5.8x more performance than the CPUs.

Only one line of legacy Fortran code had to be changed in order to achieve
the reported performance on the MIC architecture (not counting changes to the command-line interface).

The methodology discussed in this paper is directly applicable to other
bandwidth-bound stencil algorithms utilizing a hybrid OpenMP+MPI approach.


\phantom{ 
\cite{shwater}
}

\end{abstract}

\optcolsep

{\normalsize
\tableofcontents
}

\optcolend

\vfill

\shadowsize 2 pt
\centering
\fboxsep 6 pt{
\shadowbox{
\parbox{\colfaxwidth}{\footnotesize
Colfax International (\href{http://www.colfax-intl.com}{http://www.colfax-intl.com/}) is a leading provider of innovative and expertly engineered workstations, servers, clusters, storage, and personal supercomputing solutions. Colfax International is uniquely positioned to offer the broadest spectrum of high performance computing solutions, all of them completely customizable to meet your needs - far beyond anything you can get from any other name brand. Ready-to-go Colfax HPC solutions deliver significant price/performance advantages, and increased IT agility, that accelerates your business and research outcomes. Colfax International's extensive customer base includes Fortune 1000 companies, educational institutions, and government agencies. Founded in 1987, Colfax International is based in Sunnyvale, California and is privately held.
}}}

}

\opttwocolumn


\opttwocolumn

\section{Introduction}\label{sec:intro}

This paper continues our cycle of publications on the optimization of HPC applications
for computing systems enabled with Intel Xeon Phi coprocessors featuring the Many Integrated Core (MIC) architecture\footnote {See, e.g., \cite{primer} for information}.
The application studied in this work is a computational fluid dynamics code (CFD),
a shallow water equation solver. The source code of the application is freely available at \cite{shwater}.
The code is based on the numerical method developed by Robert Sadourny \cite{sadourny74} and initially written by Paul Schwarztrauber of National Center for Atmospheric Research (\href{http://www.cisl.ucar.edu/}{NCAR}).
Thereafter, the code was modernized and adapted for hybrid parallelism in the OpenMP and MPI frameworks
by Cliff Addison (University of Liverpool).
This solver is a good candidate for acceleration on the Intel MIC architecture
because it submits to parallelization and operates in the memory bandwidth-bound regime.

In our solver, just like in any MPI application designed for CPU-based clusters, 
initialization and data traffic were already implemented prior to porting to the Intel MIC architecture. 
Therefore, the porting procedure can be limited to recompilation of the code \cite{colfax2013}.
Furthermore, the solver is already optimized for compute nodes based on multi-core CPUs.
It is expressed in the Fortran language, with OpenMP and array notation used
to express parallelism.
As a consequence, the node-level performance of the solver is also very efficient
when the code is compiled for and run on an Intel Xeon Phi coprocessor.
This means that the optimization process does not need to involve any
code modification, and only the execution environment and parameters need to be tuned.

Optimization methods described here can be applied to other
bandwidth-bound stencil codes on multidimensional Cartesian grids.

In Section~\ref{sec:solver} we outline the nature of the calculation and the methodology
of work distribution across computing nodes in the cluster.
Section~\ref{sec:singlenode} discusses the optimization of the MPI environment
for executing the code on a single CPU-based node or a single Xeon Phi coprocessor.
Cluster-level tuning is discussed in Section~\ref{sec:cluster},
and results are presented in Section~\ref{sec:scalability}.

\section{Enstrophy-Conserving Shallow Water Equation Solver}\label{sec:solver}

The physical model of shallow water flow assumes a perfect incompressible fluid
subject to the gravitational force. This model can be used to describe, for example,
the movement of the ocean water on Earth
or certain atmospheric phenomena.

The quantities that describe shallow water flow are the pressure $P(x,y)$
and the fluid velocity ${\bf V}(x,y)$.
The components of vector ${\bf V}$ are denoted as $V_x\equiv u$ and $V_y \equiv v$.
These quantities can be related to the height $H(x,y)$ of the water
above the equilibrium level as $H=P+V^2/2$, where $V^2\equiv u^2 + v^2$.
The evolution of shallow water flow can be expressed with the set of equations (\ref{eq:dyn1})--(\ref{eq:dyn2}):
\begin{eqnarray}
\label{eq:dyn1}
\frac{\partial {\bf V}}{\partial t} + \eta {\bf N}\times(P{\bf V}) + \nabla \left( P + \frac12 V^2 \right) & = & 0,\\
\label{eq:dyn2}
\frac{\partial P}{\partial t} + \nabla \cdot (P {\bf V} ) & = & 0.
\end{eqnarray}
Here 
$\eta\equiv \left[ \partial v/\partial x - \partial u /\partial y \right] / P$ is the potential vorticity, and 
${\bf N}$ is the unit vector normal to the plain domain $S$.
The system of units is chosen so that the free fall acceleration is reduced to a dimensionless value $g=1$,
and the fluid density $\rho=1$.

The numerical method for solving these equations proposed by Sadourny \cite{sadourny74}
conserves enstrophy 

\begin{equation*}
Z=\frac12 \int_S \eta^2 P dS,
\end{equation*}
which is a desirable stability property. Enstrophy conservation avoids
nonlinear instability that leads to an energy catastrophe in the simulation.
The time update step in this method is expressed as 

\begin{eqnarray}
\label{eq:up1}
\frac{\partial u}{\partial t} - \left< \eta \right>_x \big< \left< V \right> _x\big>_y + \delta_x H & = & 0, \\
\label{eq:up2}
\frac{\partial v}{\partial t} + \left< \eta \right>_y \big< \left< U \right> _y\big>_x + \delta_y H & = & 0, \\
\label{eq:up3}
\frac{\partial P}{\partial t} + \delta_x U + \delta_y V & = & 0.
\end{eqnarray}
Here $\delta_x$ and $\delta_y$ are derivation operators on the staggered simulation grid (see Section 2 in \cite{sadourny74}),
triangular brackets $\left< \right>$ are operators of averaging along the direction indicated by their subscript (equivalent to
the overline operator in \cite{sadourny74}), and $U\equiv \left< P \right>_x u$ and $V\equiv \left< P \right>_y v$ are mass fluxes.
These operators use only adjacent cells for derivation and averaging, and so the algorithm can be expressed as a 2-dimensional stencil operation.

In the Fortran code expressing equations (\ref{eq:up1})--(\ref{eq:up3}), quantities $u$, $v$, $U$, $V$, $H$ and $P$ are defined
on a Cartesian grid of $N\times N$ cells, with $x$-axis being the inner matrix dimension and $y$-axis being the outer.
Periodic boundary conditions are assumed.
Parallelism is achieved by distributing the grid columns ($y$-dimension) across OpenMP threads and exposing the $x$-dimension to automatic
vectorization by the compiler. 

\begin{figure}[ht]
\includegraphics[width=\columnwidth]{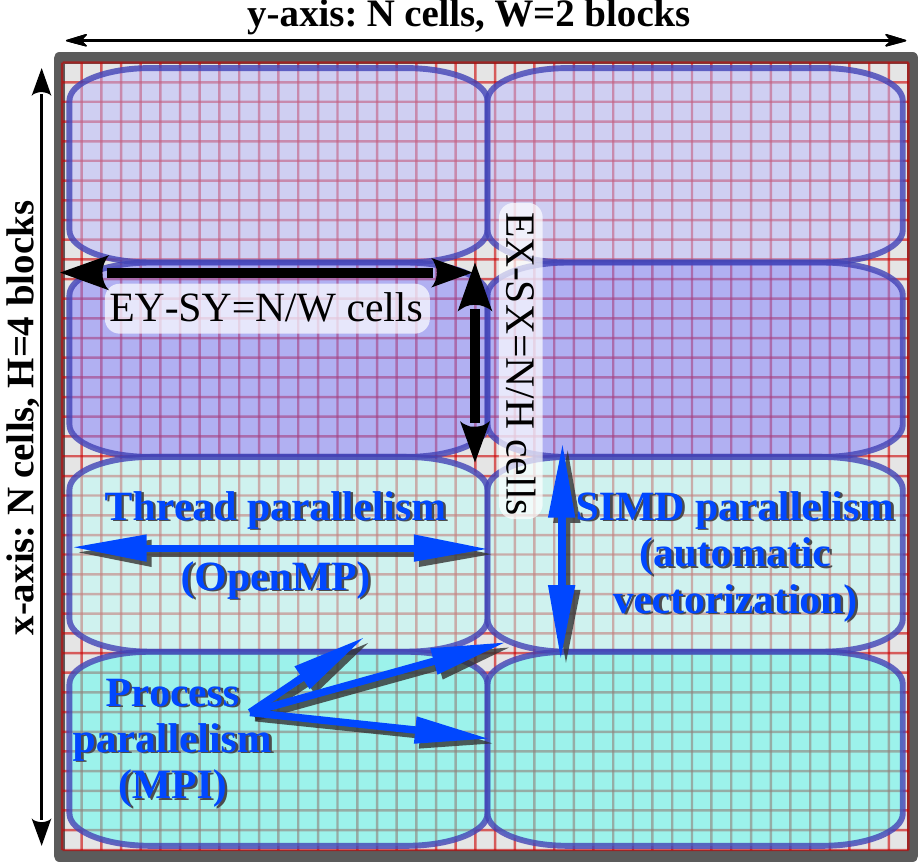}
\caption{Simulation domain of $N\times N$ cells is partitioned into $H$ blocks $x$-wise and $W$ blocks $y$-wise,
and each block is assigned to an MPI process. Within each MPI process, OpenMP
is used to distribute iterations in the $y$-dimension between threads, 
and automatic vectorization is applied in the $x$-dimension.\label{fig:gridpart}}
\end{figure}

In the distributed-memory (MPI) version of the code, the simulation grid is partitioned into $H$ blocks in the $x$-dimension $W$ in the $y$-dimension,
as shown in Figure~\ref{fig:gridpart}.
Each block is assigned to one of the $R=H\times W$ MPI processes. 
All blocks contain nearly the same number of cells.
MPI proceses exchange boundary cells in order to perform the calculation.

Listing~\ref{code:ompdo} and Figure~\ref{fig:gridpart} illustrate the approach taken in the numerical solver
to expose parallelism\footnote{Code listings in this paper are given
in the free format. The original code is expressed in the fixed-column format.}.
Listing~\ref{code:dims} shows how MPI convenience functions are used to partition the grid into blocks
for distribution across the processes of the MPI job.

\begin{listing}[ht]
\begin{Verbatim}[commandchars=\\\{\},numbers=left,firstnumber=1,stepnumber=1, ,fontsize=\scriptsize ,frame=single]
\PY{c}{!\PYZdl{}OMP PARALLEL DO schedule(static)}
\PY{k}{DO }\PY{n+nv}{J} \PY{o}{=} \PY{n+nv}{sy}\PY{p}{,}\PY{n+nv}{ey}
  \PY{c}{! Array notation below is automatically}
  \PY{c}{! vectorized by the compiler in the}
  \PY{c}{! first index dimension (x\PYZhy{}dimension).}
  \PY{n+nv}{UNEW}\PY{p}{(}\PY{n+nv}{sx}\PY{o}{+}\PY{l+m+mi}{1}\PY{p}{:}\PY{n+nv}{ex}\PY{o}{+}\PY{l+m+mi}{1}\PY{p}{,}\PY{n+nv}{J}\PY{p}{)} \PY{o}{=} \PY{n+nv}{UOLD}\PY{p}{(}\PY{n+nv}{sx}\PY{o}{+}\PY{l+m+mi}{1}\PY{p}{:}\PY{n+nv}{ex}\PY{o}{+}\PY{l+m+mi}{1}\PY{p}{,}\PY{n+nv}{J}\PY{p}{)} \PY{o}{+} \PY{p}{\PYZam{}}
    \PY{n+nv}{TDTS8}\PY{o}{*}\PY{p}{(}\PY{n+nv}{Z}\PY{p}{(}\PY{n+nv}{sx}\PY{o}{+}\PY{l+m+mi}{1}\PY{p}{:}\PY{n+nv}{ex}\PY{o}{+}\PY{l+m+mi}{1}\PY{p}{,}\PY{n+nv}{J}\PY{o}{+}\PY{l+m+mi}{1}\PY{p}{)} \PY{o}{+} \PY{n+nv}{Z}\PY{p}{(}\PY{n+nv}{sx}\PY{o}{+}\PY{l+m+mi}{1}\PY{p}{:}\PY{n+nv}{ex}\PY{o}{+}\PY{l+m+mi}{1}\PY{p}{,}\PY{n+nv}{J}\PY{p}{))}\PY{o}{*} \PY{p}{\PYZam{}}
    \PY{p}{(}\PY{n+nv}{CV}\PY{p}{(}\PY{n+nv}{sx}\PY{o}{+}\PY{l+m+mi}{1}\PY{p}{:}\PY{n+nv}{ex}\PY{o}{+}\PY{l+m+mi}{1}\PY{p}{,}\PY{n+nv}{J}\PY{o}{+}\PY{l+m+mi}{1}\PY{p}{)} \PY{o}{+} \PY{n+nv}{CV}\PY{p}{(}\PY{n+nv}{sx}\PY{p}{:}\PY{n+nv}{ex}\PY{p}{,}\PY{n+nv}{J}\PY{o}{+}\PY{l+m+mi}{1}\PY{p}{)} \PY{o}{+} \PY{p}{\PYZam{}}
     \PY{n+nv}{CV}\PY{p}{(}\PY{n+nv}{sx}\PY{p}{:}\PY{n+nv}{ex}\PY{p}{,}\PY{n+nv}{J}\PY{p}{)} \PY{o}{+} \PY{n+nv}{CV}\PY{p}{(}\PY{n+nv}{sx}\PY{o}{+}\PY{l+m+mi}{1}\PY{p}{:}\PY{n+nv}{ex}\PY{o}{+}\PY{l+m+mi}{1}\PY{p}{,}\PY{n+nv}{J}\PY{p}{))} \PY{o}{\PYZhy{}} \PY{p}{\PYZam{}}
    \PY{n+nv}{TDTSDX}\PY{o}{*}\PY{p}{(}\PY{n+nv}{H}\PY{p}{(}\PY{n+nv}{sx}\PY{o}{+}\PY{l+m+mi}{1}\PY{p}{:}\PY{n+nv}{ex}\PY{o}{+}\PY{l+m+mi}{1}\PY{p}{,}\PY{n+nv}{J}\PY{p}{)} \PY{o}{\PYZhy{}} \PY{n+nv}{H}\PY{p}{(}\PY{n+nv}{sx}\PY{p}{:}\PY{n+nv}{ex}\PY{p}{,}\PY{n+nv}{J}\PY{p}{))}
  \PY{n+nv}{VNEW}\PY{p}{(}\PY{n+nv}{sx}\PY{o}{+}\PY{l+m+mi}{1}\PY{p}{:}\PY{n+nv}{ex}\PY{o}{+}\PY{l+m+mi}{1}\PY{p}{,}\PY{n+nv}{J}\PY{p}{)} \PY{o}{=} \PY{n+nv}{VOLD}\PY{p}{(}\PY{n+nv}{sx}\PY{o}{+}\PY{l+m+mi}{1}\PY{p}{:}\PY{n+nv}{ex}\PY{o}{+}\PY{l+m+mi}{1}\PY{p}{,}\PY{n+nv}{J}\PY{p}{)} \PY{o}{+} \PY{p}{\PYZam{}}
    \PY{n+nv}{TDTS8}\PY{o}{*}\PY{p}{(}\PY{n+nv}{Z}\PY{p}{(}\PY{n+nv}{sx}\PY{o}{+}\PY{l+m+mi}{1}\PY{p}{:}\PY{n+nv}{ex}\PY{o}{+}\PY{l+m+mi}{1}\PY{p}{,}\PY{n+nv}{J}\PY{o}{+}\PY{l+m+mi}{1}\PY{p}{)} \PY{o}{+} \PY{n+nv}{Z}\PY{p}{(}\PY{n+nv}{sx}\PY{p}{:}\PY{n+nv}{ex}\PY{p}{,}\PY{n+nv}{J}\PY{o}{+}\PY{l+m+mi}{1}\PY{p}{))}\PY{o}{*} \PY{p}{\PYZam{}}
    \PY{p}{(}\PY{n+nv}{CU}\PY{p}{(}\PY{n+nv}{sx}\PY{o}{+}\PY{l+m+mi}{1}\PY{p}{:}\PY{n+nv}{ex}\PY{o}{+}\PY{l+m+mi}{1}\PY{p}{,}\PY{n+nv}{J}\PY{o}{+}\PY{l+m+mi}{1}\PY{p}{)} \PY{o}{+} \PY{n+nv}{CU}\PY{p}{(}\PY{n+nv}{sx}\PY{p}{:}\PY{n+nv}{ex}\PY{p}{,}\PY{n+nv}{J}\PY{o}{+}\PY{l+m+mi}{1}\PY{p}{)} \PY{o}{+} \PY{p}{\PYZam{}}
     \PY{n+nv}{CU}\PY{p}{(}\PY{n+nv}{sx}\PY{p}{:}\PY{n+nv}{ex}\PY{p}{,}\PY{n+nv}{J}\PY{p}{)} \PY{o}{+} \PY{n+nv}{CU}\PY{p}{(}\PY{n+nv}{sx}\PY{o}{+}\PY{l+m+mi}{1}\PY{p}{:}\PY{n+nv}{ex}\PY{o}{+}\PY{l+m+mi}{1}\PY{p}{,}\PY{n+nv}{J}\PY{p}{))} \PY{o}{\PYZhy{}} \PY{p}{\PYZam{}}
    \PY{n+nv}{TDTSDY}\PY{o}{*}\PY{p}{(}\PY{n+nv}{H}\PY{p}{(}\PY{n+nv}{sx}\PY{p}{:}\PY{n+nv}{ex}\PY{p}{,}\PY{n+nv}{J}\PY{o}{+}\PY{l+m+mi}{1}\PY{p}{)} \PY{o}{\PYZhy{}} \PY{n+nv}{H}\PY{p}{(}\PY{n+nv}{sx}\PY{p}{:}\PY{n+nv}{ex}\PY{p}{,}\PY{n+nv}{J}\PY{p}{))}
  \PY{n+nv}{PNEW}\PY{p}{(}\PY{n+nv}{sx}\PY{p}{:}\PY{n+nv}{ex}\PY{p}{,}\PY{n+nv}{J}\PY{p}{)} \PY{o}{=} \PY{n+nv}{POLD}\PY{p}{(}\PY{n+nv}{sx}\PY{p}{:}\PY{n+nv}{ex}\PY{p}{,}\PY{n+nv}{J}\PY{p}{)} \PY{o}{\PYZhy{}} \PY{p}{\PYZam{}}
    \PY{n+nv}{TDTSDX}\PY{o}{*}\PY{p}{(}\PY{n+nv}{CU}\PY{p}{(}\PY{n+nv}{sx}\PY{o}{+}\PY{l+m+mi}{1}\PY{p}{:}\PY{n+nv}{ex}\PY{o}{+}\PY{l+m+mi}{1}\PY{p}{,}\PY{n+nv}{J}\PY{p}{)} \PY{o}{\PYZhy{}} \PY{n+nv}{CU}\PY{p}{(}\PY{n+nv}{sx}\PY{p}{:}\PY{n+nv}{ex}\PY{p}{,}\PY{n+nv}{J}\PY{p}{))} \PY{o}{\PYZhy{}}
    \PY{n+nv}{TDTSDY}\PY{o}{*}\PY{p}{(}\PY{n+nv}{CV}\PY{p}{(}\PY{n+nv}{sx}\PY{p}{:}\PY{n+nv}{ex}\PY{p}{,}\PY{n+nv}{J}\PY{o}{+}\PY{l+m+mi}{1}\PY{p}{)} \PY{o}{\PYZhy{}} \PY{n+nv}{CV}\PY{p}{(}\PY{n+nv}{sx}\PY{p}{:}\PY{n+nv}{ex}\PY{p}{,}\PY{n+nv}{J}\PY{p}{))}
\PY{k}{END DO}
\PY{c}{!\PYZdl{}OMP END PARALLEL DO}
\end{Verbatim}
\vskip -0.5em
\vskip -0.1in
\caption{Time update step in the solver in Fortran with OpenMP and array syntax. 
Iterations along the $y$-dimension are distributed across OpenMP threads,
and iterations along the $x$-dimension can be automatically vectorized by the compiler.
The length of the inner loop, $ex-sx+1$, depends on the problem size and on the partitioning
of the problem between MPI processes.
The corresponding numerical model is expressed by Equations~(\ref{eq:up1})--(\ref{eq:up3}).\label{code:ompdo}}
\end{listing}

\begin{listing}[hbt]
\begin{Verbatim}[commandchars=\\\{\},numbers=left,firstnumber=1,stepnumber=1, ,fontsize=\scriptsize ,frame=single]
\PY{n+nv}{dims}\PY{p}{(}\PY{l+m+mi}{1}\PY{p}{)} \PY{o}{=} \PY{l+m+mi}{0}
\PY{n+nv}{dims}\PY{p}{(}\PY{l+m+mi}{2}\PY{p}{)} \PY{o}{=} \PY{l+m+mi}{0}
\PY{k}{call }\PY{n+nv}{MPI\PYZus{}DIMS\PYZus{}CREATE}\PY{p}{(} \PY{n+nv}{numprocs}\PY{p}{,} \PY{l+m+mi}{2}\PY{p}{,} \PY{n+nv}{dims}\PY{p}{,} \PY{n+nv}{ierr} \PY{p}{)}
\PY{k}{call }\PY{n+nv}{MPI\PYZus{}CART\PYZus{}CREATE}\PY{p}{(} \PY{n+nv}{MPI\PYZus{}COMM\PYZus{}WORLD}\PY{p}{,} \PY{l+m+mi}{2}\PY{p}{,} \PY{n+nv}{dims}\PY{p}{,} \PY{p}{\PYZam{}}
     \PY{n+nv}{periods}\PY{p}{,} \PY{n+nb}{.false.}\PY{p}{,} \PY{n+nv}{comm2d}\PY{p}{,} \PY{n+nv}{ierr} \PY{p}{)}
\PY{k}{call }\PY{n+nv}{MPI\PYZus{}CART\PYZus{}GET}\PY{p}{(} \PY{n+nv}{comm2d}\PY{p}{,} \PY{l+m+mi}{2}\PY{p}{,} \PY{n+nv}{dims}\PY{p}{,} \PY{n+nv}{periods}\PY{p}{,} \PY{p}{\PYZam{}}
     \PY{n+nv}{coords}\PY{p}{,} \PY{n+nv}{ierr} \PY{p}{)}
\PY{k}{call }\PY{n+nv}{MPE\PYZus{}DECOMP1D}\PY{p}{(}\PY{n+nv}{m}\PY{p}{,} \PY{n+nv}{dims}\PY{p}{(}\PY{l+m+mi}{1}\PY{p}{),} \PY{n+nv}{coords}\PY{p}{(}\PY{l+m+mi}{1}\PY{p}{),} \PY{n+nv}{sx}\PY{p}{,} \PY{n+nv}{ex}\PY{p}{)}
\PY{k}{call }\PY{n+nv}{MPE\PYZus{}DECOMP1D}\PY{p}{(}\PY{n+nv}{n}\PY{p}{,} \PY{n+nv}{dims}\PY{p}{(}\PY{l+m+mi}{2}\PY{p}{),} \PY{n+nv}{coords}\PY{p}{(}\PY{l+m+mi}{2}\PY{p}{),} \PY{n+nv}{sy}\PY{p}{,} \PY{n+nv}{ey}\PY{p}{)}
\end{Verbatim}
\vskip -0.5em
\caption{Partitioning the simulation domain between MPI processes using MPI convenience functions.\label{code:dims}}
\end{listing}

\section{Single-Node Performance Optimization}\label{sec:singlenode}

We run all benchmarks on a cluster with the configuration summarized below:
\begin{enumerate}[1)]
\tighterenum
\item{Two compute nodes \href{http://www.colfax-intl.com/nd/workstations/sxp8600p.aspx}{Colfax ProEdge\texttrademark\ SXP8600p} interconnected with Gigabit Ethernet interconnects and with \href{http://www.mellanox.com/page/products_dyn?product_family=61}{Mellanox MHQH19B-XTR} HCAs (QDR, ConnectX-2 VPI), one per node;}
\item{Each compute node contains a dual-socket \mbox{\href{http://ark.intel.com/products/75283/Intel-Xeon-Processor-E5-2697-v2-30M-Cache-2_70-GHz}{Intel E5-2697 v2}} processor (12 cores per socket, Ivy Bridge architecture) with 128~GB of DDR3 RAM at 1600~MHz;}
\item{In each node, four 61-core \href{http://ark.intel.com/products/75799/Intel-Xeon-Phi-Coprocessor-7120P-16GB-1_238-GHz-61-core}{Intel Xeon Phi 7120P} coprocessors are installed;}
\item{The software configuration includes the \mbox{CentOS~6.5} Linux operating system with kernel version \mbox{2.6.32-431.el6.x86\_64}, Intel Cluster Studio~XE~2013~SP1 and Intel~MPSS~3.2.1 with the OFED-1.5.4.1 stack.}
\end{enumerate}
For consistency of benchmarks, all power management features are disabled on Intel Xeon Phi coprocessors.

In the following discussion, we use the traditional terminology in the field of computing accelerators.
By CPU, processor, host, or host system, we generally mean the Intel Xeon CPU-based computing system.
By coprocessor, device or accelerator we mean the Intel Xeon Phi coprocessor.
We will also use the term ``compute device'' in contexts where the term applies to both the coprocessor and the CPU.

\begin{figure}[ht]
\includegraphics[width=\columnwidth]{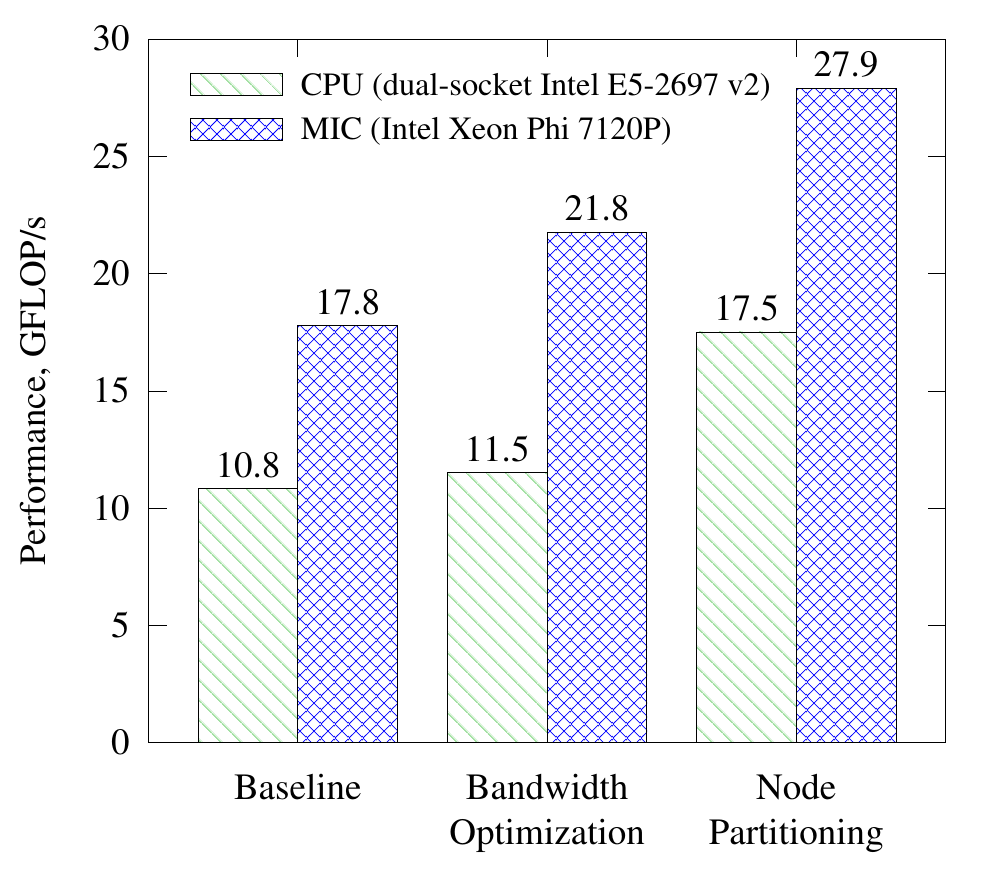}
\caption{Single-node performance: dual-socket Intel Xeon E5-2697v2 processor and Intel Xeon Phi 7120P coprocessor.
See Section~\ref{sec:singlenode} for discussion.\label{fig:singlenode}}
\end{figure}

\subsection{Baseline}

Without any platform-specific configuration of the execution environment,
the code can be run on the CPUs of the cluster compute nodes,
as well as on the Intel Xeon Phi coprocessors.
We use \texttt{mpirun} to start the calculation with a single MPI process
first on the dual-socket CPU, then on one of the Xeon Phi coprocessors, as shown in Listing~\ref{lst:base}.
By default, the hybrid OpenMP/MPI code spawns as many OpenMP threads as there are
logical cores on the respective device: 48 threads on the CPU and 244 threads 
on the coprocessor.

\begin{listing}
\begin{Verbatim}[commandchars=\\\{\}, ,fontsize=\scriptsize ,frame=single]
\PY{g+gp}{cfx@c001\PYZhy{}n001\PYZdl{}} mpirun \PYZhy{}host c001\PYZhy{}n001 \PY{l+s+se}{\PYZbs{}}
\PY{g+gp}{\PYZgt{}} \PYZhy{}np 1 /home/cfx/shallow/ncar\PYZhy{}solver1 10000
\PY{g+go}{ ...}
\PY{g+go}{ WALL CLOCK TIME FOR JOB =   29.97804 seconds}
\PY{g+go}{ EXPECTED GFLOPS RATE    =    10.84127}

\PY{g+gp}{cfx@c001\PYZhy{}n001\PYZdl{}} export \PY{n+nv}{I\PYZus{}MPI\PYZus{}MIC}\PY{o}{=}1
\PY{g+gp}{cfx@c001\PYZhy{}n001\PYZdl{}} mpirun \PYZhy{}host c001\PYZhy{}n001\PYZhy{}mic0 \PY{l+s+se}{\PYZbs{}}
\PY{g+gp}{\PYZgt{}} \PYZhy{}np 1 /home/cfx/shallow/ncar\PYZhy{}solver1.MIC 10000
\PY{g+go}{ ...}
\PY{g+go}{ WALL CLOCK TIME FOR JOB =   18.28000 seconds}
\PY{g+go}{ EXPECTED GFLOPS RATE    =    17.77900}
\end{Verbatim}
\vskip -0.5em
\caption{Baseline of code performance: execution on a single node (CPU and MIC) without tuning.\label{lst:base}}
\end{listing}

The attained performance is measured internally in the code 
by dividing the number of arithmetic operations involved in the calculation
by the elapsed wall clock time.
The result is reported in terms of GFLOP/s, which is a metric specific
to this particular solver.
For the baseline test, the CPU attained 10.8~GFLOP/s and the coprocessor 17.8~GFLOP/s.
This corresponds to the first set of bars in Figure~\ref{fig:singlenode}.

\subsection{Optimizing MPI Process Pinning for Memory Bandwidth}\label{sec:pinning}

Most modern Intel processor architectures support hyper-threading,
which is hardware support for operating more than one software thread per physical core.
While hyper-threading improves the performance of applications bound by memory access latency,
it is counter-productive for bandwidth-bound workloads that stream contiguous data from memory to cores.

In most cases (see, e.g., \cite{cr-bandwidth}), a bandwidth-bound application performs optimally on the Intel architecture
when both of the following conditions are met:
\begin{enumerate}[1)]
\tighterenum
\item one software thread is used per core, and
\item thread migration from core to core is forbidden at the operating system level.
\end{enumerate}

Therefore, the first step towards improving the performance of our shallow water equation solver
is limiting the number of OpenMP threads on the coprocessor, and
enforcing the pinning of each thread to a respective physical core.
This is done by passing the environment variable \texttt{OMP\_NUM\_THREADS=24}
to the MPI process running on the host, or \texttt{OMP\_NUM\_THREADS=61} to the MPI process on the coprocessor,
as shown in Listing~\ref{lst:bandwidth}.

\begin{listing}
\begin{Verbatim}[commandchars=\\\{\}, ,fontsize=\scriptsize ,frame=single]
\PY{g+gp}{cfx@c001\PYZhy{}n001\PYZdl{}} mpirun \PYZhy{}host c001\PYZhy{}n001 \PY{l+s+se}{\PYZbs{}}
\PY{g+gp}{\PYZgt{}} \PYZhy{}np 1 \PYZhy{}env \PY{l+s+s2}{\PYZdq{}OMP\PYZus{}NUM\PYZus{}THREADS=24\PYZdq{}} \PY{l+s+se}{\PYZbs{}}
\PY{g+gp}{\PYZgt{}} /home/cfx/shallow/ncar\PYZhy{}solver1 10000
\PY{g+go}{ ...}
\PY{g+go}{ WALL CLOCK TIME FOR JOB =   28.23723 seconds}
\PY{g+go}{ EXPECTED GFLOPS RATE    =    11.50963}

\PY{g+gp}{cfx@c001\PYZhy{}n001\PYZdl{}} export \PY{n+nv}{I\PYZus{}MPI\PYZus{}MIC}\PY{o}{=}1
\PY{g+gp}{cfx@c001\PYZhy{}n001\PYZdl{}} mpirun \PYZhy{}host c001\PYZhy{}n001\PYZhy{}mic0 \PY{l+s+se}{\PYZbs{}}
\PY{g+gp}{\PYZgt{}} \PYZhy{}np 1  \PYZhy{}env \PY{l+s+s2}{\PYZdq{}OMP\PYZus{}NUM\PYZus{}THREADS=61\PYZdq{}} \PY{l+s+se}{\PYZbs{}}
\PY{g+gp}{\PYZgt{}} /home/cfx/shallow/ncar\PYZhy{}solver1.MIC 10000
\PY{g+go}{ ...}
\PY{g+go}{ WALL CLOCK TIME FOR JOB =   14.93185 seconds}
\PY{g+go}{ EXPECTED GFLOPS RATE    =    21.76555}
\end{Verbatim}
\vskip -0.5em
\caption{Memory bandwidth is optimized by setting the number of OpenMP threads to the number of physical cores. Affinity
of threads and processes to specific cores is set by the Intel MPI library automatically.\label{lst:bandwidth}}
\end{listing} 

Bandwidth optimization by restricting the number of threads
improves performance on the CPU by 6\%, and on the coprocessor by 22\%
compared to the baseline.
This is shown by the second set of bars in Figure~\ref{fig:singlenode}.

Note that without MPI, the optimization of OpenMP environment for bandwidth-bound
applications requires setting \texttt{OMP\_NUM\_THREADS} to the number of physical cores,
and then setting \texttt{KMP\_AFFINITY=scatter} to prevent thread migration across cores.
With MPI, the latter step is not necessary, as the pinning functionality of Intel MPI
takes care of binding threads to cores.
Pinning decisions made by Intel MPI can be diagnosed by setting \texttt{I\_MPI\_DEBUG=4}
prior to running a job. Environment variables for pinning control
are described in the \href{https://software.intel.com/sites/products/documentation/hpc/ics/impi/41/lin/Reference_Manual/hh_goto.htm#Interoperability_with_OpenMP.htm}{Intel MPI Library Reference Manual} \cite{impi}.

\subsection{Node Partitioning}\label{sec:nodepart}

Further improvement of the shallow water equation solver
may come from using multiple MPI processes per device
with proportionally fewer OpenMP threads per process.
This tweak in the configuration improves performance 
for two reasons.

\begin{enumerate}[1)]
\tighterenum
\item On the host system, the two CPU sockets form a Non-Uniform Memory Access (NUMA) system,
in which each CPU socket can access memory attached to it faster than memory attached to the other CPU.
This leads to performance penalties when one MPI process with multiple threads operates on a large
data set, and thread affinity to data is not set. On the other hand, with two (or, in general, with an even number of)
MPI processes on a dual-socket system, pinning will ensure that each process addresses only its local memory. 

\item On both the CPU and the coprocessor, partitioning the dataset leads to better cache traffic
in the stencil operation expressed by the code in Listing~\ref{code:ompdo}. Indeed,
the inner loop expressed by array notation has \texttt{ex-sx+1} iterations.
This value is equal to the grid size $N$ for only one MPI process,
but for a job with several MPI processes distributed across the simulation box,
the value of \texttt{ex-sx+1} will be smaller.
Smaller length of the inner loop improves data reuse.
Indeed, arrays \texttt{Z} and \texttt{H} used in the calculation of \texttt{UNEW}
are subsequently used to compute \texttt{VNEW}; and arrays \texttt{CU} and \texttt{CV}
re-used in the calculation of \texttt{PNEW}.
Furthermore, because the stencil operation uses adjacent \texttt{y}-values,
some arrays are re-used in subsequent \texttt{J}-iterations.
For a small enough value of \texttt{ex-sx+1}, the re-used arrays
may still be in the processor's cache when they are accessed a second time.
Thus, reducing \texttt{ex-sx+1} by increasing the number of MPI processes per compute device improves
cache utilization.

\end{enumerate}

The phenomenon described above is often exploited in shared-memory 
applications (e.g., \cite{cr-transposition2}) through an optimization known as loop tiling, or loop blocking.
Loop tiling improves performance by increasing the temporal locality of data access.
Our solver would benefit from loop tiling, however, it is not implemented.
At the same time, a similar pattern of data re-use occurs naturally
when OpenMP is used in tandem with MPI to partition the simulation box.

\begin{figure}[ht]
\includegraphics[width=\columnwidth]{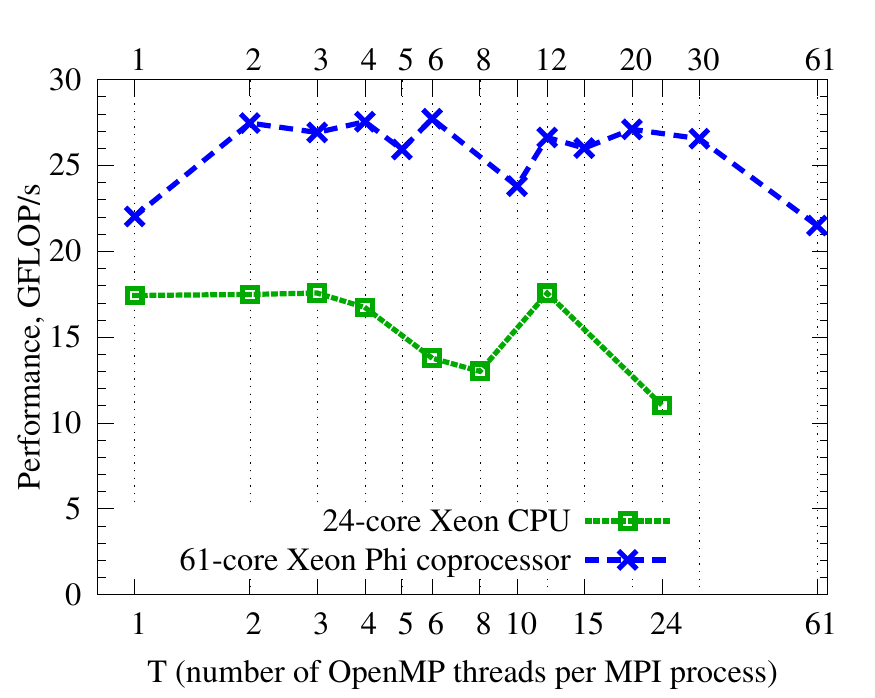}
\caption{Impact of node partitioning between several MPI processes on performance.
Data obtained on the host and on the coprocessor.
In each run, the number of MPI processes is chosen as $24/T$
on the CPU, and as $61/T$ on the coprocessor (with rounding). \label{fig:threads}}
\end{figure}

Figure~\ref{fig:threads} demonstrates the dependence of the
code performance on the number of OpenMP threads per MPI process.
On the host CPU, using fewer threads and many processes leads to better performance
than using only one 24-threaded process.
On the coprocessor, exploratory trials are needed to determine the
optimal number of threads. 
The case of $T=2$ in Figure~\ref{fig:threads} corresponds to the third set of bars
in Figure~\ref{fig:singlenode}.

\section{Cluster-Level Optimization Caveats}\label{sec:cluster}

At this point, the performance of the solver on a single node is tuned,
and one Xeon Phi coprocessor performs 1.6x faster than the dual-socket Xeon host.
Considering that each of our compute nodes contains 4 coprocessors,
we can expect to run up 6.4x faster on the MIC accelerators than
on the CPUs. Because in this case, the performance of Xeon Phi coprocessors
dwarfs the performance of host CPUs, 
from this point on, we will focus only on optimization for coprocessors.

\begin{figure}[htb]
\includegraphics[width=\columnwidth]{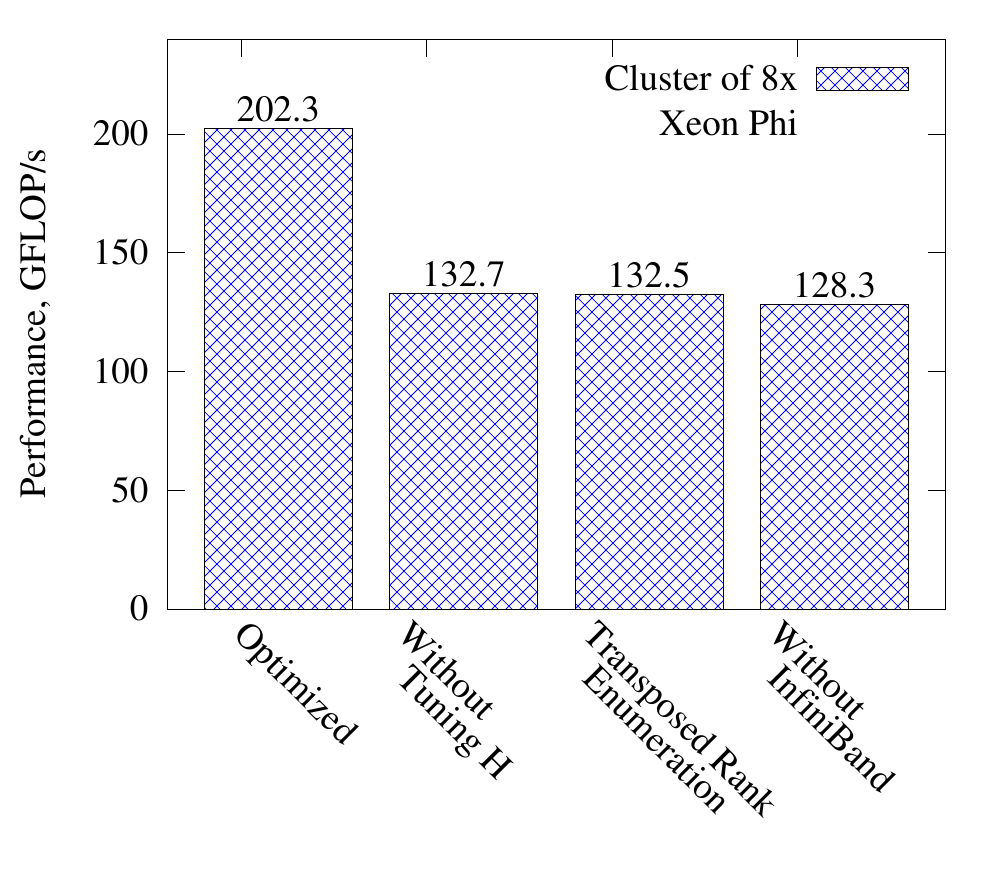}
\caption{Tuned performance and degraded performance of eight Xeon Phi coprocessors
in two compute nodes of the cluster. See Section~\ref{sec:cluster} for discussion.\label{fig:caveats}}
\end{figure}

\subsection{Domain Decomposition Tuning}

Scaling across multiple compute nodes with MPI can be done
by modifying the arguments of \texttt{mpirun} and using a machine file
as illustrated in Listing~\ref{lst:multi}.
There we configure MPI to run the code on 8 coprocessors
installed in two compute nodes: \texttt{c001-n001} and \texttt{c001-n002}.
We use 4 threads per process and 15 processes per Xeon Phi
(a total of $15\times 8=120$ MPI ranks).
This is one of the optimal numbers of threads as shown 
in Figure~\ref{fig:threads}.
We could not use 2 threads per process, because the job would exceed
the maximum number of ranks supported in our cluster.

\begin{listing}
\begin{Verbatim}[commandchars=\\\{\}, ,fontsize=\scriptsize ,frame=single]
\PY{g+gp}{cfx@c001\PYZhy{}n001\PYZdl{}} cat allmics.txt
\PY{g+go}{c001\PYZhy{}n001\PYZhy{}mic0:15}
\PY{g+go}{c001\PYZhy{}n001\PYZhy{}mic1:15}
\PY{g+go}{c001\PYZhy{}n001\PYZhy{}mic2:15}
\PY{g+go}{c001\PYZhy{}n001\PYZhy{}mic3:15}
\PY{g+go}{c001\PYZhy{}n002\PYZhy{}mic0:15}
\PY{g+go}{c001\PYZhy{}n002\PYZhy{}mic1:15}
\PY{g+go}{c001\PYZhy{}n002\PYZhy{}mic2:15}
\PY{g+go}{c001\PYZhy{}n002\PYZhy{}mic3:15}
\PY{g+gp}{cfx@c001\PYZhy{}n001\PYZdl{}} export \PY{n+nv}{I\PYZus{}MPI\PYZus{}MIC}\PY{o}{=}1
\PY{g+gp}{cfx@c001\PYZhy{}n001\PYZdl{}} mpirun  \PY{l+s+se}{\PYZbs{}}
\PY{g+gp}{\PYZgt{}} \PYZhy{}machine allmics.txt \PYZhy{}env \PY{l+s+s2}{\PYZdq{}OMP\PYZus{}NUM\PYZus{}THREADS=4\PYZdq{}}\PY{l+s+se}{\PYZbs{}}
\PY{g+gp}{\PYZgt{}} /home/cfx/shallow/ncar\PYZhy{}solver1.MIC 10000
\PY{g+go}{ ...}
\PY{g+go}{ WALL CLOCK TIME FOR JOB =    2.44866 seconds}
\PY{g+go}{ EXPECTED GFLOPS RATE    =   132.72564}
\end{Verbatim}
\vskip -0.5em
\caption{Running the solver on a cluster of
2 compute nodes with 4 coprocessors each,
6 threads per process and 10 processes per coprocessor.\label{lst:multi}}
\end{listing}

As Listing~\ref{lst:multi} reports, we obtain a performance of 132.7~GFLOP/s
from 8 coprocessors, which is 4.8x greater than the performance of one
coprocessor. This is not a satisfactory scalability metric, and more investigation
is required.

Going back to Figure~\ref{fig:gridpart}, we recognize that with a greater number
of MPI ranks, the simulation grid is partitioned into smaller blocks.
This drives the calculation away from the optimal value of the x-loop length \texttt{ex-sx+1}
that was achieved by tuning the number of threads $T$ on one compute device (see Section~\ref{sec:nodepart}).
Consequently, one needs to tune the number of threads per MPI rank, $T$,
for the multi-node run, rather than rely on the tuned value of $T$ for a single-node run.

Furthermore, one can infer from Listing~\ref{code:dims} that the code allows MPI
to choose the decomposition of the simulation domain into blocks both in the $x$-dimension
(parameter $H$) and in the $y$-dimension (parameter $W$). Considering how important the value of \texttt{ex-sx+1} 
is for performance on Xeon Phi coprocessors (see Figure~\ref{fig:threads}),
we should consider tuning the dimensions of the blocks $H$ and $W$.
This can be done by changing the call to \texttt{MPI\_DIMS} as shown in
Listing~\ref{code:newdims}.

\begin{listing}[bht]
\begin{Verbatim}[commandchars=\\\{\},numbers=left,firstnumber=1,stepnumber=1, ,fontsize=\scriptsize ,frame=single]
\PY{c}{! Value of H taken from command\PYZhy{}line arguments}
\PY{n+nv}{dims}\PY{p}{(}\PY{l+m+mi}{1}\PY{p}{)} \PY{o}{=} \PY{n+nv}{H}
\PY{n+nv}{dims}\PY{p}{(}\PY{l+m+mi}{2}\PY{p}{)} \PY{o}{=} \PY{l+m+mi}{0}
\PY{k}{call }\PY{n+nv}{MPI\PYZus{}DIMS\PYZus{}CREATE}\PY{p}{(} \PY{n+nv}{numprocs}\PY{p}{,} \PY{l+m+mi}{2}\PY{p}{,} \PY{n+nv}{dims}\PY{p}{,} \PY{n+nv}{ierr} \PY{p}{)}
\PY{c}{! ...}
\end{Verbatim}
\vskip -0.5em
\caption{Adjusting the decomposition of simulation grid
into blocks by setting a value of $H$ (number of blocks in the
$x$-dimension), instead of allowing MPI to choose a value.
MPI will choose the second decomposition parameter, $W$, so that $H\times W=R$ ($R$ is the total number of MPI processes).
This is the only code change that was necessary for optimization on the MIC architecture.
Compare to Listing~\ref{code:dims}.
\label{code:newdims}}
\end{listing}

Of course, the total number of blocks $W\times H$ must be equal to the
number of MPI ranks in the simulation, $R$ (otherwise, 
the call to \texttt{MPI\_DIMS} will fail). 
Therefore, we 
should only probe values of $H$ that are divisors of $R$.
The value of $R$ is computed as $R=(60/T)*M$, where $T$ is the number of
threads per rank, and $M$ is the number of coprocessors in the calculation.

\begin{figure}[H]
\includegraphics[width=\columnwidth]{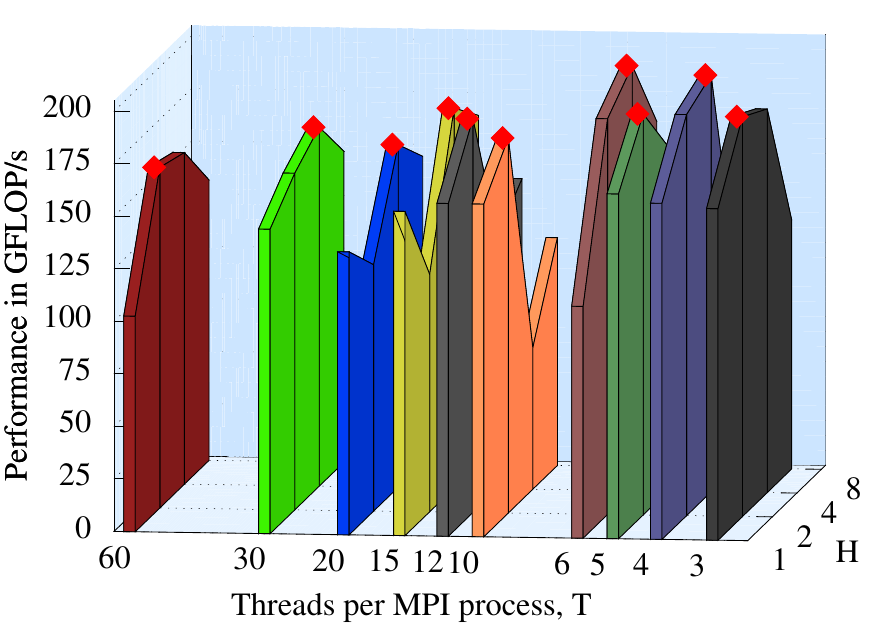}
\caption{Two parameters must be empirically determined for each problem size $N$
and cluster size $M$: (i) $T$, the number of threads per process, and (ii) $H$, the
number of blocks in the $x$-dimension for grid partitioning.
The best-performing pair of $(T,H)$ can then be used in production runs.
Shown are results of calibration for $N=10000$, $M=8$. Red markers
indicate the best values of $H$ for a given $T$.\label{fig:tuning}}
\end{figure}

A thorough scan of parameter space $(T,H)$ allows to tune our solver
to the cluster size on which it is run. Figure~\ref{fig:tuning}
shows the performance as a function of these two parameters.
For $M=8$ coprocessors, we restricted the search to values
of $T\in {1,2,3,4,5,6,10,12,15,20,30,60}$ and $H\in{1,2,4,8}$.

The best combination of the tuning parameters for problem size $N=10000$
on $M=8$ coprocessors is $T=6$, $H=4$ (see Table~\ref{tab:speedup}).
Compared to similarly tuned performance of the solver on one coprocessor,
this is a 7.2x speedup. 
This performance is reflected in the first bar of Figure~\ref{fig:caveats},
while the performance before tuning $T$ and $H$ is the second bar in this Figure.

\subsection{MPI Rank Numbering}\label{sec:numbering}

Even though at this point, we have provided an optimized
solution to scaling the shallow water equation solver,
we will demonstrate some caveats on cluster-level tuning,
based our experience in this project.

In Listing~\ref{lst:multi} we used a machine file in which
each line has format \texttt{host:ppn}, where \texttt{host}
is the hostname of a coprocessor, and \texttt{ppn} is the number
of MPI processes on that compute device. In this case,
MPI rank numbers are assigned to each coprocessor
in contiguous chunks.
In Listing~\ref{lst:nontransposed} we illustrate
rank numbering for 6 MPI processes
on 2 coprocessors.

\begin{listing}[tp]
\begin{Verbatim}[commandchars=\\\{\}, ,fontsize=\scriptsize ,frame=single]
\PY{g+gp}{cfxuser@c001\PYZhy{}n001\PYZdl{}} export \PY{n+nv}{I\PYZus{}MPI\PYZus{}DEBUG}\PY{o}{=}3
\PY{g+gp}{cfxuser@c001\PYZhy{}n001\PYZdl{}} cat mics1.txt
\PY{g+go}{mic0:3}
\PY{g+go}{mic1:3}
\PY{g+gp}{[cfxuser@c001\PYZhy{}n001\PYZdl{}} mpirun \PY{l+s+se}{\PYZbs{}}
\PY{g+gp}{\PYZgt{}} \PYZhy{}machine mics1.txt \PYZhy{}env \PY{l+s+s2}{\PYZdq{}OMP\PYZus{}NUM\PYZus{}THREADS=20\PYZdq{}} \PY{l+s+se}{\PYZbs{}}
\PY{g+gp}{\PYZgt{}} /home/cfx/shallow/ncar\PYZhy{}solver1.MIC 10000
\PY{g+go}{...}
\PY{g+go}{[0] MPI startup(): Rank  Pid     Node name}
\PY{g+go}{[0] MPI startup(): 0     20647   c001\PYZhy{}n001\PYZhy{}mic0}
\PY{g+go}{[0] MPI startup(): 1     20648   c001\PYZhy{}n001\PYZhy{}mic0}
\PY{g+go}{[0] MPI startup(): 2     20649   c001\PYZhy{}n001\PYZhy{}mic0}
\PY{g+go}{[0] MPI startup(): 3     15250   c001\PYZhy{}n001\PYZhy{}mic1}
\PY{g+go}{[0] MPI startup(): 4     15251   c001\PYZhy{}n001\PYZhy{}mic1}
\PY{g+go}{[0] MPI startup(): 5     15252   c001\PYZhy{}n001\PYZhy{}mic1}
\PY{g+go}{...}
\end{Verbatim}
\vskip -0.5em
\caption{Specifying the number of processes
per node in the machine file assignes a contiguous
chunk of ranks to each compute device.\label{lst:nontransposed}}
\end{listing}

At the same time, it is possible to start a calculation
in a different way, as shown 
in Listing~\ref{lst:transposed}.
In that case, we list one compute device per line in the machine file,
and add the argument \texttt{-np=R} to \texttt{mpirun},
where \texttt{R} is the number of MPI ranks equal to 6 in this case.
As apparent from the listing, now compute devices are assigned to ranks
in the round-robin order.

\begin{listing}[htp]
\begin{Verbatim}[commandchars=\\\{\}, ,fontsize=\scriptsize ,frame=single]
\PY{g+gp}{cfxuser@c001\PYZhy{}n001\PYZdl{}} export \PY{n+nv}{I\PYZus{}MPI\PYZus{}DEBUG}\PY{o}{=}3
\PY{g+gp}{cfxuser@c001\PYZhy{}n001\PYZdl{}} cat mics2.txt
\PY{g+go}{mic0}
\PY{g+go}{mic1}
\PY{g+gp}{cfxuser@c001\PYZhy{}n001\PYZdl{}} mpirun \PYZhy{}np 6 \PY{l+s+se}{\PYZbs{}}
\PY{g+gp}{\PYZgt{}} \PYZhy{}machine mics1.txt \PYZhy{}env \PY{l+s+s2}{\PYZdq{}OMP\PYZus{}NUM\PYZus{}THREADS=20\PYZdq{}} \PY{l+s+se}{\PYZbs{}}
\PY{g+gp}{\PYZgt{}} /home/cfx/shallow/ncar\PYZhy{}solver1.MIC 10000
\PY{g+go}{...}
\PY{g+go}{[0] MPI startup(): Rank  Pid     Node name}
\PY{g+go}{[0] MPI startup(): 0     20809   c001\PYZhy{}n001\PYZhy{}mic0}
\PY{g+go}{[0] MPI startup(): 1     15413   c001\PYZhy{}n001\PYZhy{}mic1}
\PY{g+go}{[0] MPI startup(): 2     20810   c001\PYZhy{}n001\PYZhy{}mic0}
\PY{g+go}{[0] MPI startup(): 3     15414   c001\PYZhy{}n001\PYZhy{}mic1}
\PY{g+go}{[0] MPI startup(): 4     20811   c001\PYZhy{}n001\PYZhy{}mic0}
\PY{g+go}{[0] MPI startup(): 5     15415   c001\PYZhy{}n001\PYZhy{}mic1}
\PY{g+go}{...}
\end{Verbatim}
\vskip -0.5em
\caption{Specifying the total number of MPI processes
using the argument \texttt{-np}, and listing compute nodes
in the machine file assigns nodes to ranks
in the round-robin order.\label{lst:transposed}}
\end{listing}

\begin{figure}[ht]
\centering
\includegraphics[width=0.45\columnwidth]{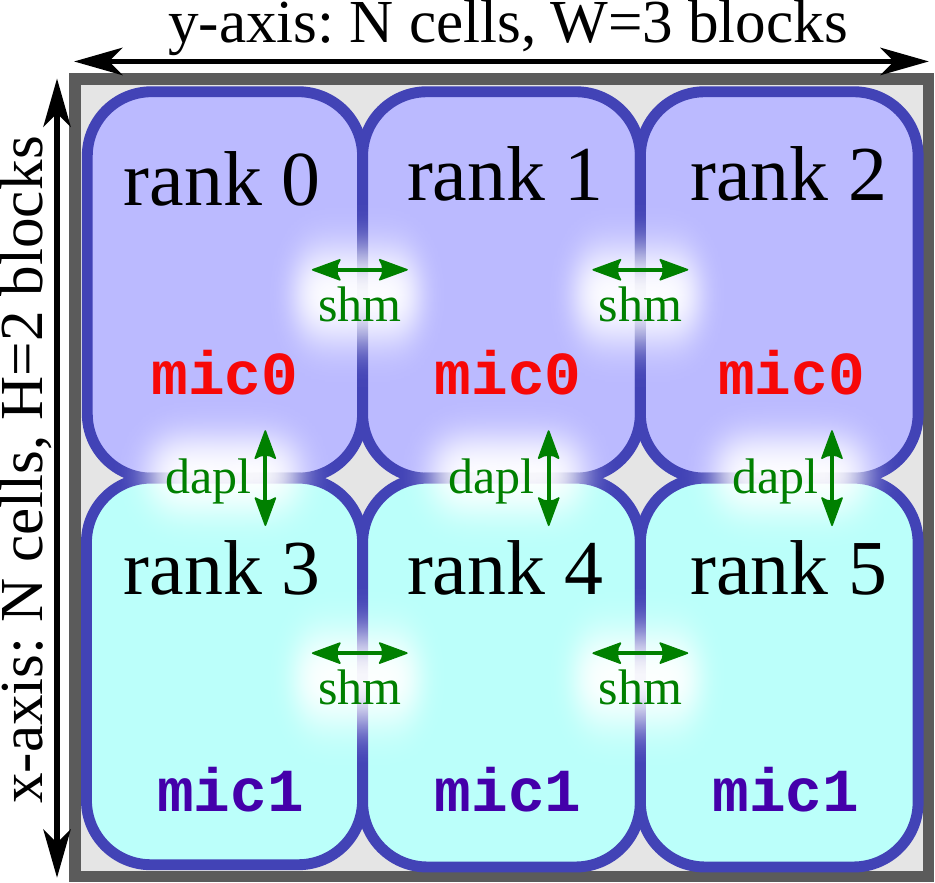}
\hskip 0.10in
\includegraphics[width=0.45\columnwidth]{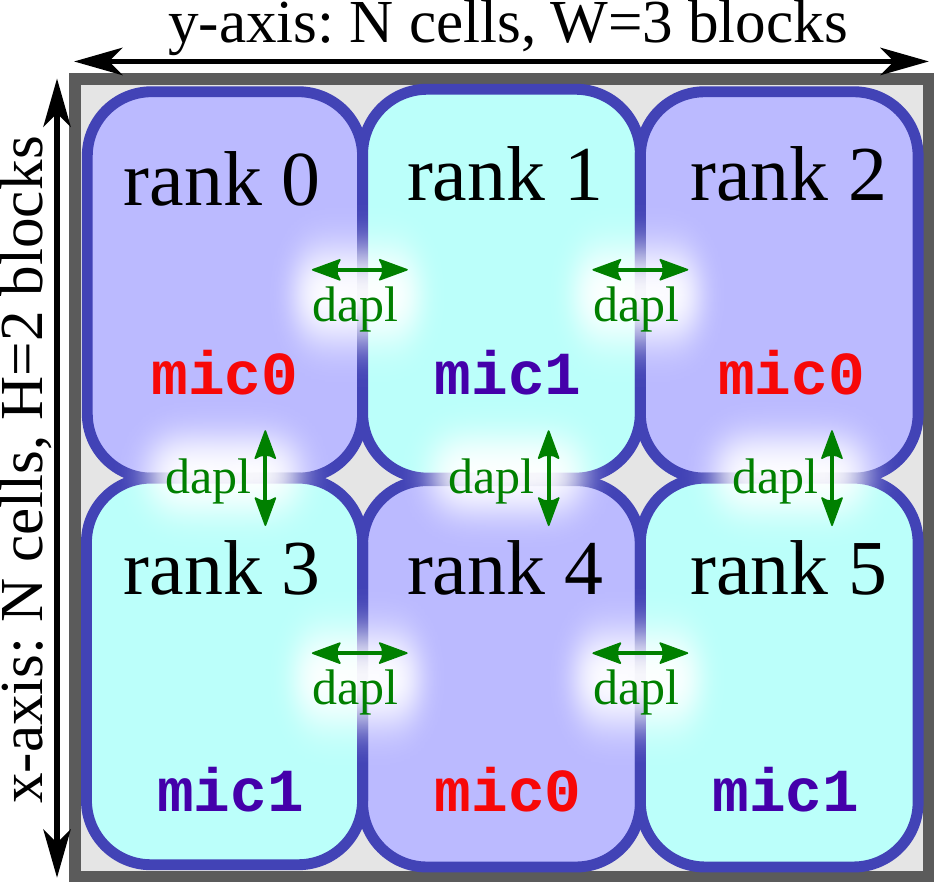}
\caption{Rank assignment in a grid of $R=W \times H$ processes.
{\em Left:} optimized rank enumeration with the \texttt{:ppn} suffix in the machine file (Listing~\ref{lst:nontransposed}). 
Most left and right neighbors are local (on the same coprocessor) and communicate via the shared-memory copy protocol \texttt{shm}.
{\em Right:} sub-optimal rank enumeration with the \texttt{-np=R} argument of \texttt{mpirun} (Listing~\ref{lst:transposed}). 
All left/right and top/bottom neighbors are remote, communicating via the InfiniBand protocol \texttt{dapl}.
\label{fig:ranks}}
\end{figure}

Figure~\ref{fig:ranks} illustrates the rank assignment pattern in these two cases.
In our solver, MPI communication is set up in such a way that each process
communicates only with the processes that operate on adjacent grid blocks
(\mbox{left/right} and \mbox{top/bottom} neighbors).
Consequently, when the left and right neighbors are local (i.e., on the same coprocessor) rather than remote (i.e., on a different device), MPI communication will take less time. 
That is because intra-coprocessor communication will be performed using the shared-memory copy protocol \texttt{shm},
relieving the strain on the PCIe and InfiniBand subsystems that carry inter-coprocessor traffic with the protocol \texttt{dapl}
(see \cite{cr-ib} for details).
Thus, the configuration on the left-hand side of Figure~\ref{fig:ranks} (with host file in format \texttt{host:ppn})
is better optimized than the configuration on the right-hand side (with argument \texttt{-np=R}).

Indeed, a benchmark shows that running our shallow water equation solver
on a grid of processors transposed in this way yields only 132.5~GFLOP/s
on eight compute devices (third bar in Figure~\ref{fig:caveats}), as opposed to 202.3~GFLOP/s in the optimized case.

\subsection{Role of Communication Fabric}\label{sec:fabric}

Our compute cluster uses InfiniBand interconnects for MPI communication.
InfiniBand drivers are integrated with Intel MPSS in order to enable 
Coprocessor Communication Link (CCL). 
This functionality virtualizes an InfiniBand HCA on each coprocessor,
which enables peer-to-peer messaging between Intel Xeon Phi coprocessors in different
compute nodes, and also improves comunication between coprocessors
within the same host.
See \cite{cr-ib} for more details.

Without InfiniBand adapters, and without the specialized OFED stack,
the cluster would be using
the TCP protocol over Gigabit Ethernet for communication.
We can emulate this situation by setting the environment variable
\texttt{I\_MPI\_FABRICS=tcp} for the calculation,
which makes the Intel MPI library use the TCP protocol
over the Gigabit Ethernet fabric.

The fourth bar in Figure~\ref{fig:caveats} shows the performance
of the tuned calculation on 8 coprocessors in 2 compute nodes
with the TCP fabric. It achieves 128.3~GFLOP/s, which is only 63\%
of the performance with InfiniBand.

\section{Parallel Scalability}\label{sec:scalability}

Scaling the shallow water equation solver across multiple compute nodes
requires that for each cluster size $M$ and problem size $N$, the tuning parameters $(T,H)$
are empirically determined in a calibration run. $T$ is the number
of threads per MPI process, and $H$ is the number of
blocks in the $x$-dimension for partitioning the simulation domain
between MPI processes. 
$T$ must be a divisor of 60 (otherwise,
several Xeon Phi cores may go unused), 
and $H$ must be a divisor of the
number of processors $R=(60/T)*M$. We have also determined
that values of $H$ greater than $M$ are inefficient, due to 
the nature of the communication pattern discussed in Section~\ref{sec:numbering}. 

Considering these pruning rules, there
are usually several tens of pairs $(T,H)$ that must be tested.
Calibration runs take several seconds each, and therefore
the process of tuning can be realistically completed on a time
scale of several minutes.

\begin{figure}[ht]
\includegraphics[width=\columnwidth]{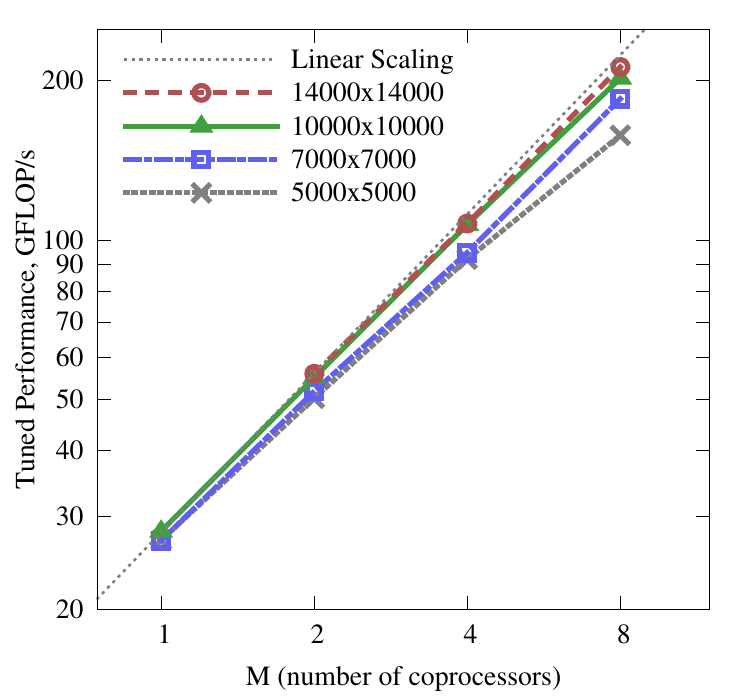}
\caption{Parallel scalability with tuning of the parameter set $(T,H)$ (see Table~\ref{tab:speedup}) 
of the shallow water equation solver on a cluster of Intel Xeon Phi coprocessors
for different problem sizes.\label{fig:scal}}
\end{figure}

Table~\ref{tab:speedup} shows the results obtained 
for different problem sizes: $N\in{5000, 7000, 10000, 14000}$.
For $N=10000$, the working dataset is around 9~GB,
so this is the biggest problem in the studied set that
fits into the 16~GB memory of a 7120P Xeon Phi coprocessor.
Figure~\ref{fig:scal} illustrates the best performance
as a function of the number of coprocessors.

\begin{table}[ht]
\footnotesize
\centering
\begin{tabular}{llllll}
\hline\hline
Size N & MICs & Threads & H & GFLOP/s & Speedup \\ 
\hline 
5000 & 1 & 6 & 2 & 27.1 & 1.00 \\ 
5000 & 2 & 6 & 1 & 50.2 & 1.85 \\ 
5000 & 4 & 6 & 1 & 92.0 & 3.40 \\ 
5000 & 8 & 6 & 2 & 157.8 & 5.82 \\ 
\hline
7000 & 1 & 15 & 2 & 27.0 & 1.00 \\ 
7000 & 2 & 10 & 3 & 51.8 & 1.92 \\ 
7000 & 4 & 12 & 4 & 94.7 & 3.51 \\ 
7000 & 8 & 6 & 2 & 185.1 & 6.85 \\ 
\hline
10000 & 1 & 2 & 3 & 28.1 & 1.00 \\ 
10000 & 2 & 2 & 3 & 55.0 & 1.96 \\ 
10000 & 4 & 3 & 4 & 107.0 & 3.81 \\ 
10000 & 8 & 6 & 4 & 202.3 & 7.20 \\ 
\hline
14000 & 1 & n/a & n/a & n/a & n/a \\
14000 & 2 & 3 & 5 & 55.9 & 2.00$^*$ \\ 
14000 & 4 & 6 & 5 & 107.4 & 3.84$^*$ \\ 
14000 & 8 & 4 & 4 & 212.4 & 7.60$^*$ \\ 
\hline
\hline
\end{tabular}
\caption{Optimized performance and values tuning parameters $T$ (threads per MPI process) and $H$ (grid blocks in $x$-dimension) for problem sizes $N\in{5000,7000,10000,14000}$
on $M=1$, $2$, $4$ and $8$ Intel Xeon Phi coprocessors. \\
$^*$ For $N=14000$, the problem does not fit in memory of a single coprocessor, and 
reported speedup is normalized so that the performance for $M=2$ has a speedup of 2.0.\label{tab:speedup}}
\end{table}

We have found that the speedup approaches linear scaling law
for larger problems, which is a common behavior in massively parallel
systems. For our specific application, this is explained 
by the fact that the amount of calculation scales with the problem
size as $O(N^2)$, while the amount of communication scales as $O(N)$.
At the same time, problems that are large enough to demonstrate
perfect scalability on 8 coprocessors do not fit into the memory
of one coprocessor.

With calculations running only on CPUs, the performance is not as sensitive
to the values tuning parameters. Scalability is easy to achieve with CPUs, especially
because in our two-node cluster, we only have to scale across two CPU
architecture devices as opposed to eight MIC accelerators.

\begin{figure}[ht]
\includegraphics[width=\columnwidth]{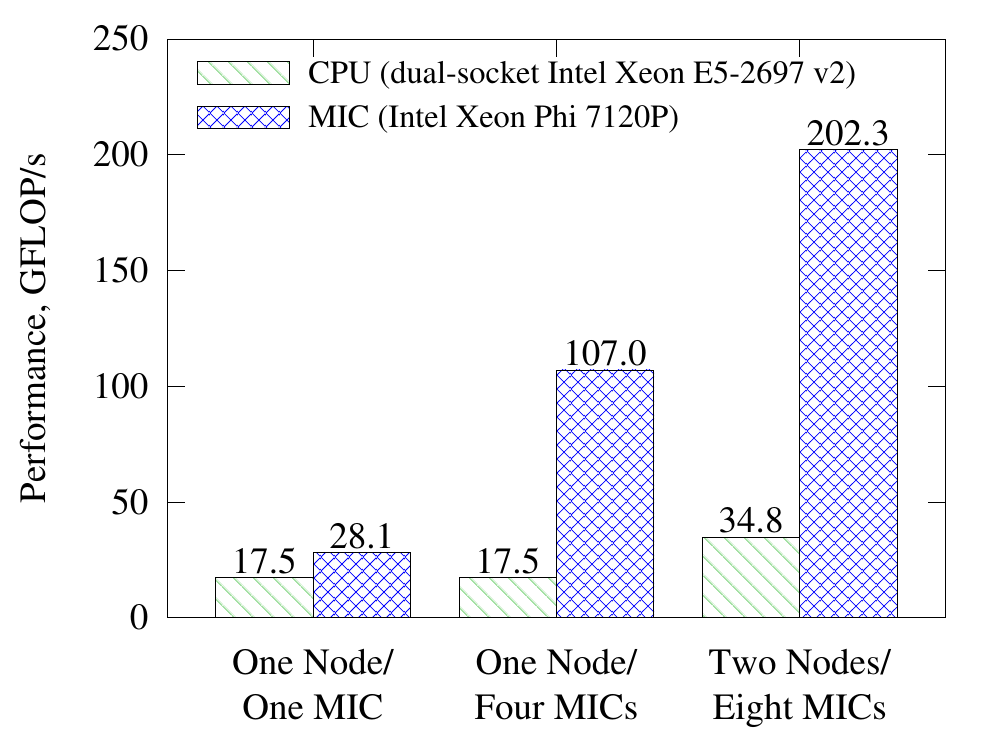}
\caption{Scalability across multiple coprocessors and multiple CPUs. 
In this plot, the second and third set of bars show the performance achieved 
within the same rack space with only CPUs and only MIC accelerators (one node in the second set of bars and two nodes in the third set).
\label{fig:clustr}}
\end{figure}

The tuned performance results on both platforms are shown in Figure~\ref{fig:clustr}.
In this plot, the second set of bars ``Two Nodes/Eight MICs'' shows the performance
achieved within the same rack space (two nodes) with only CPUs and only MIC accelerators.

\section{Conclusions}

We analyzed the performance of a shallow water equation solver
on a MIC-enabled computing cluster.
The numerical method is a 2-dimensional stencil code
operating in the memory bandwidth-bound regime.

We demonstrated how tuning the number of threads per process $T$
improves performance by increasing the re-use of cached data.
A second tuning parameter is required for optimal performance with Intel Xeon Phi coprocessor.
This parameter, $H$, tweaks the size of the inner loops by changing the number
of partitions of the simulation domain.

In the parameter calibration process, it was apparent that Xeon Phi coprocessors
are less forgiving of sub-optimal parameters than Xeon CPUs.
However, after tuning, we achieved a 7.2x performance increase with eight coprocessors
compared to one (202.3~GFLOP/s on eight Xeon Phi 7120P). On two compute nodes with dual-socket E5-2697v2 CPUs
we achieved 34.8~GFLOP/s.
This amounts to a $202.3/34.8=5.8$x speedup from the MIC architecture over using the CPUs of two compute nodes alone.

Our optimizations for the MIC architecture required a modification of only one line of code
(not counting modifications of the code interface), which controls
the partitioning of the simulation domain between MPI ranks (see Listing~\ref{code:newdims}). 
The original code in Fortran, optimized for CPU-based clusters,
performs well as long as the MPI environment was optimized.

It is important to note that the code was not specifically tuned for Intel Xeon Phi coprocessors prior to this work.
In fact, the last updates to the code relevant to performance date to the year 2003, when Intel MIC architecture was not in existence.
This confirms once again the notions that
\begin{enumerate}[i)]
\tighterenum 
\item an application that efficiently uses multi-core CPUs is likely to also perform well on Intel Xeon Phi coprocessors without code modification,
\item Intel MIC architecture is a viable platform for acceleration and modernization of time-tested legacy applications.
\end{enumerate}


\small

\onecolumn

\end{document}